\documentclass[12pt]{article}
\usepackage{epsfig}
\usepackage{amsmath}
\usepackage{amssymb}
\usepackage{cite}

\date{\today}

\newcommand{\be}{\begin{equation}}
\newcommand{\ee}{\end{equation}}
\newcommand{\bea}{\begin{eqnarray}}
\newcommand{\eea}{\end{eqnarray}}
\newcommand{\bml}{\begin{mathletters}}
\newcommand{\eml}{\end{mathletters}}

\begin{document}
\thispagestyle{empty}
\vspace*{0.2cm}
\begin{center}
{\Large \bf On the classicality of bosonic stars}
\\
\vspace{1.0cm}
{\large Carlos A. R. Herdeiro\footnote{herdeiro@ua.pt; corresponding author}
 and Eugen Radu\footnote{eugen.radu@ua.pt}} 
\\
\vspace{0.5cm}
\small{
 Departamento de Matem\'atica da Universidade de Aveiro and \\
Center for Research and Development in Mathematics and Applications (CIDMA), 
  \\
  Campus de Santiago, 3810-183 Aveiro, Portugal} 
\end{center}

\begin{abstract} 
Because the Klein-Gordon and Proca equations involve $\hbar$, they describe quantum fields. Their solutions, however, may be treated as classical if their typical action obeys $S^{\rm typical}\gg \hbar$. This is possible due to their bosonic nature, allowing states with many particles. We show, by generic arguments, that the typical action for such bosonic stars is ${\mathcal{S}^{\rm typical}}/{\hbar}\gtrsim  \left({M^{\rm max}}/{M_{\rm Pl}}\right)^2\sim 10^{76}\left({M^{\rm max}}/{M_\odot}\right)^2$, 
where 
$M^{\rm max}$ is the maximal bosonic star mass of the particular model and 
$M_{\rm Pl}$ is the Planck mass. 
Thus, for models allowing $M^{\rm max}\gg M_{\rm Pl}$ and for solutions with mass $\sim M^{\rm max}$, the classical treatment is legitimate, which  includes masses in the astrophysical interesting range $\gtrsim M_\odot$.
\end{abstract}


\vspace*{1.cm}

\begin{center}
{\footnotesize{Essay written for the Gravity Research Foundation 2022 Awards for Essays on Gravitation.}}
\\
{\footnotesize{Submitted March 28th, 2022}}

\end{center}
\newpage
\vspace{0.5cm}



\section{Introduction }

Scalar boson stars are solutions of the Einstein--Klein-Gordon (EKG) equations, first discussed in the late 1960s~\cite{Kaup:1968zz,Ruffini:1969qy}. More recently, a cousin model with a great deal of similarities was unveiled in the form of vector boson stars ($aka$ Proca stars)~\cite{Brito:2015pxa}, which are solutions of the Einstein--Proca (EP) equations.  Collectively we shall be calling scalar and vector boson stars as \textit{bosonic stars}.

Over the last decades bosonic stars have increasingly been considered as possible astrophysical, macroscopic objects~\cite{Schunck:2003kk,Liebling:2012fv}, playing roles as putative dark matter constituents or black hole mimickers - see $e.g.$~\cite{Bustillo:2020syj,Herdeiro:2021lwl}. In particular, in the latter context bosonic stars should have a mass $M_{\rm BS}$ in the astrophysical mass range of black holes, roughly
\begin{equation}
M_{\rm BS} \ \in \ [1,10^{10}] \, M_\odot \ .
\label{massinterval}
\end{equation}

 Within such astrophysical discussions, bosonic stars are treated as \textit{classical} objects.  Yet, the Klein-Gordon and Proca equations are quantum, relativistic equations. This leads to the often raised question whether a classical description of bosonic stars is acceptable. The purpose of this essay is to (quantitatively) clarify this question.

The key point is the basic fact of \textit{scale hierarchy} in physics. Dimensionful scales often determine regions of the parameter space where distinct physical treatments and the use of convenient approximations are possible. That is, some simplified model can be used up (or down) to some scale, but ignored physical effects become important beyond that scale, invalidating the approximation. For instance, a non-relativistic treatment is a good approximation if the typical velocities $v^{\rm typical}$ in a given system are well below that of light $c$, $i.e.$, $v^{\rm typical}\ll c$. Analogously, quantum effects can be ignored if the typical action $\mathcal{S}^{\rm typical}$ of a system is much greater than the reduced Planck's constant $\hbar$, $i.e.$ $\mathcal{S}^{\rm typical}\gg \hbar$.

In the following we shall assess whether bosonic stars have a typical action much larger than Planck's constant. We shall keep all basic constants $(c,\hbar, G)$ in the discussion below, for clarity.

\section{The bosonic field theories }

\subsection{Klein-Gordon}
The Klein-Gordon equation~\cite{Klein:1926tv,Gordon} is motivated by considering the relativistic energy-momentum relation $E^2-{\bf p}^2c^2=m^2c^4$,
where $m$ is the rest mass of a particle, $E$ its energy and ${\bf p}$ its 4-momentum. Associating to the energy and 4-momentum the operators $E \rightarrow i\hbar {\partial}/{\partial t}$ and  ${\bf p} \rightarrow -i\hbar \nabla$, 
the mass-shell relation becomes an operatorial relation acting on a field $\Phi(x^\alpha)$, where $x^\alpha=(ct,x^i)$, $i=1,2,3$:
\begin{equation}
\Box \Phi=\frac{m^2c^2}{\hbar^2} \Phi \ ,
\label{KG}
\end{equation}
where $\Box$ is the (Minkowski spacetime) d'Alembertian operator. 
Equation~\eqref{KG} is the Klein-Gordon equation, and the explicit appearence of $\hbar$ and $c$ manifest its relativistic quantum nature. The scalar field can be taken as real or complex. Choosing the latter, the Klein-Gordon equation can be derived from the action
\begin{equation}
\mathcal{S}^{\Phi}=\hbar \int d^4 x\sqrt{-g}\left[-g^{\alpha\beta}\partial_\alpha\Phi^*\partial_\beta\Phi -\mu^2\Phi\Phi^* \right] \ ,
\label{kgaction}
\end{equation}
where $^*$ denotes complex conjugation, 
\begin{equation}
\mu\equiv \frac{mc}{\hbar} \ 
\label{icw}
\end{equation}
is the \textit{inverse} Compton wavelength of the field and the generic metric $g_{\alpha\beta}$ with determinant $g$ makes the action covariant. The choice of the constant in front of the action~\eqref{kgaction} determines the dimension of the scalar field $\Phi$. The choice used ($i.e.$, $\hbar$) implies
\begin{equation}
[\Phi]=\frac{1}{L} \ ,
\end{equation}
where $L$ stands for length. This choice makes the quantum nature of the scalar field action manifest.

The scalar field theory described by the action~\eqref{kgaction} is invariant under a global $U(1)$ symmetry, that transforms the field by a global phase,  $\Phi\rightarrow  e^{i\chi}  \Phi$, where $\chi$ is constant. This leads to a conserved (in the sense of a local continuity equation) Noether current
\begin{equation}
j^\alpha=-i(\Phi^*\partial^\alpha \Phi-\Phi \partial^\alpha \Phi^*) \ .
\end{equation}
Clearly $D_\alpha j^\alpha=0$, where $D_\alpha$ is the covariant derivative of $g_{\alpha\beta}$. Then, the integral on a spacelike slice $\Sigma$ of the temporal component of the current is a Noether charge, denoted $Q$,
\begin{equation}
Q\equiv \int_{\Sigma} j^t \ .
\label{nc}
\end{equation}
If follows that $Q$ is dimensionless. It is interpreted as the particle number, 
an interpretation made precise upon quantization~\cite{Itzykson:1980rh}.

\subsubsection{The typical action for scalar boson stars}
\label{section3}
%
The scalar field described by the action~\eqref{kgaction} can be minimally coupled to gravity, by considering the EKG action,
\begin{equation}
\mathcal{S}^{\rm EKG}=\frac{c^3}{16\pi G}\int d^4x \sqrt{-g} R +\hbar \int d^4 x\sqrt{-g}\left[-g^{\alpha\beta}\partial_\alpha\Phi^*\partial_\beta\Phi -\mu^2\Phi\Phi^* \right]\ ,
\label{ekgaction}
\end{equation}
where $R$ is the Ricci scalar of $g_{\alpha\beta}$.
The EKG equations are
\begin{equation}
G_{\alpha\beta}=\frac{8\pi G}{c^4} T_{\alpha\beta} \ , \qquad D_\alpha\partial^\alpha\Phi=\mu^2\Phi \ ,
\label{ekg}
\end{equation}
where the Klein-Gordon energy-momentum tensor is
\begin{equation}
T_{\alpha\beta}=\hbar c\left\{\partial_\alpha\Phi^*\partial_\beta \Phi +\partial_\alpha\Phi\partial_\beta \Phi^*-g_{\alpha\beta}\left[ g^{\sigma\tau}\partial_\sigma\Phi^*\partial_\tau \Phi+\mu^2\Phi^*\Phi   \right]  \right\} \ .
\end{equation}
One can check, $e.g.$, that $T_{00}$ is, in terms of dimensions, an energy density, $[T_{00}]=M/L^3$.

The simplest scalar boson star solutions to the coupled EKG equations~\eqref{ekg} are obtained in spherical symmetry~\cite{Kaup:1968zz,Ruffini:1969qy}. This can be achieved under the metric ansatz
\begin{equation}
ds^2=-N(r)\sigma^2(r) c^2dt^2+\frac{dr^2}{N(r)}+r^2(d\theta^2+\sin^2\theta d\varphi^2) \ , \qquad N(r)\equiv 1-\frac{2m(r)}{r} \ ,
\label{metric}
\end{equation}
and scalar field ansatz
\begin{equation}
\Phi=\phi(r) e^{-i\omega t} \ ,
\end{equation}
which introduces the oscillating frequency of the field,  $\omega>0$.
The profile functions have the following dimensions:
\begin{equation}
 [\sigma(r)]=L^0  \ , \qquad [m(r)]=L \ , \qquad  [\phi(r)]=\frac{1}{L} \ . 
\end{equation}
These functions must be obtained numerically - see $e.g.$~\cite{Schunck:2003kk}. The Noether charge~\eqref{nc} can be expressed in terms of these:
\begin{equation}
Q
= \frac{8\pi \omega}{c}\int dr \frac{r^2\phi(r)^2}{N(r)\sigma(r)} \ .
\label{ncbs}
\end{equation}

We can now use~\eqref{ncbs} to express the typical action for a scalar boson star solution. To estimate the typical action we consider a convenient term in the action~\eqref{ekgaction}:
\begin{equation}
\mathcal{S}^{\rm typical}=\hbar\int d^4x r^2\sin\theta \sigma(r)\left[-g^{00}\partial_0\Phi^*\partial_0\Phi\right]=  \frac{1}{2}\hbar \omega \, \Delta t \, Q \ , 
\label{tabs}
\end{equation}
where $\Delta t$ denotes a typical time scale, which can be factorized since the Lagrangian density becomes time independent for the boson star solutions.

The result~\eqref{tabs} is rather intuitive: the typical action is the fundamental state energy of a single harmonic oscillator ($\hbar \omega/2$), times the number of particles $Q$, multiplied by the time scale considered for the observation or physical process considered, $\Delta t$. This corroborates the  interpretation that scalar boson stars are $Q$ scalar quanta in the same (fundamental) state.

\subsection{Proca}
The Proca equations~\cite{Proca:1936fbw} are a generalization of the relativistic Maxwell equations by the addition of a mass term. Taking such mass, $m$, as the one of the fundamental quantum of the field, the Proca equations are as quantum as the Klein-Gordon equation. This is unlike the Maxwell's equations, which albeit quantizable, make no mention to the properties of individual photons. 

The (covariant) Proca equations read
\begin{equation}
D_{\alpha}\mathcal{F}^{\alpha\beta}=\mu^2\mathcal{A}^\beta \ ,
\label{peq}
\end{equation}
where $\mathcal{A}_\beta$ stands for the co-variant components of the Proca potential, whereas the corresponding field strength is $\mathcal{F}_{\alpha\beta}=\partial_\alpha\mathcal{A}_\beta-\partial_\beta\mathcal{A}_\alpha$. As before, $\mu$ is the inverse Compton wavelength of the Proca field, which relates to the physical mass of the particles $m$ via~\eqref{icw}.

As in the Klein-Gordon case, the Proca field can be taken as real or complex. Choosing the latter, the Proca equations~\eqref{peq} can be derived from the action 
\begin{equation}
\mathcal{S}^{\Phi}=\hbar \int d^4 x\sqrt{-g}\left[-\frac{1}{4}\mathcal{F}^{\alpha\beta}\mathcal{F}_{\alpha\beta}^* -\frac{\mu^2}{2}\mathcal{A}^{\alpha}\mathcal{A}_{\alpha}^* \right] \ .
\label{paction}
\end{equation}
The choice of the constant in front of the Proca action~\eqref{paction} determines the dimension of the Proca field $\mathcal{A}_\alpha$. The choice used ($i.e.$, $\hbar$) implies
\begin{equation}
[\mathcal{A}_\alpha]=\frac{1}{L} \ .
\end{equation}
Again, this choice makes the quantum nature of the Proca field action manifest.

The complex Proca field theory described by the action~\eqref{paction} is invariant under a global $U(1)$ symmetry, that transforms the field by a global phase,  $\mathcal{A}_\beta\rightarrow  e^{i\chi}  \mathcal{A}_\beta$, where $\chi$ is constant. The conserved  Noether current now reads
\begin{equation}
j^\alpha=\frac{i}{2}\left\{(\mathcal{F}^*)^{\alpha\beta}\mathcal{A}_\beta -\mathcal{F}^{\alpha\beta}\mathcal{A}_\beta^*\right\} \ .
\end{equation}
Clearly $D_\alpha j^\alpha=0$. The corresponding Noether charge, again denoted $Q$, is also computed by eq.~\eqref{nc}; it is also dimensionless and interpreted as the particle number.

\subsubsection{The typical action for Proca stars}
Minimally coupling the Proca field described by the action~\eqref{paction} to gravity yields the EP action,
\begin{equation}
\mathcal{S}^{\rm EP}=\frac{c^3}{16\pi G}\int d^4x \sqrt{-g} R +\hbar \int d^4 x\sqrt{-g}\left[-\frac{1}{4}\mathcal{F}^{\alpha\beta}\mathcal{F}_{\alpha\beta}^* -\frac{\mu^2}{2}\mathcal{A}^{\alpha}\mathcal{A}_{\alpha}^* \right] \ .
\label{Paction}
\end{equation}
The field equations are the Einstein equations in~\eqref{ekg} together with Proca equations~\eqref{peq}. The Proca energy-momentum tensor is
\begin{equation}
T_{\alpha\beta}=\hbar c\left\{-\mathcal{F}_{\sigma(\alpha}^*{\mathcal{F}}_{\beta)}^{\ \, \sigma}-\frac{1}{4}g_{\alpha\beta}\mathcal{F}_{\sigma\tau}^*{\mathcal{F}}^{\sigma\tau} +\mu^2\left[\mathcal{A}_{(\alpha}{\mathcal{A}}^*_{\beta)}-\frac{1}{2}g_{\alpha\beta}\mathcal{A}_\sigma^*{\mathcal{A}}^\sigma \right]\right\}  \ .
\end{equation}

To obtain the simplest Proca star solutions, which are spherical, the metric ansatz is still given by~\eqref{metric}, 
whereas the Proca field ansatz is~\cite{Brito:2015pxa}
\begin{equation}
\mathcal{A}=e^{-i\omega t}\left[F(r)\, c dt+iG(r) \, dr\right] \ .
\label{ansatzp}
\end{equation}
The Noether charge can again be obtained in terms of the unknown metric and Proca profile functions
\begin{equation}
Q=4\pi \int dr \frac{r^2 G(r)}{\sigma(r)} \left[\frac{\omega}{c}G(r)-F'(r)\right] \ ,
\end{equation}
where prime denotes radial derivative.  
To estimate the typical action we again consider a set of  convenient terms in the action~\eqref{Paction}:
\begin{equation}
\mathcal{S}^{\rm typical}=\hbar\int d^4x r^2\sin\theta \sigma(r)\left[-\frac{1}{4}\mathcal{F}^{\alpha\beta}\mathcal{F}_{\alpha\beta}^* -\frac{\mu^2}{2}g^{00}\mathcal{A}_{0}\mathcal{A}_{0}^*\Phi\right]=  \frac{1}{2}\hbar \omega \, \Delta t \, Q \ , 
\label{taps}
\end{equation}
where the field equations were used in the last equality. We thus recover precisely the same intuitive form for the typical action as in the scalar case.

\section{The Noether charge of an astrophysical bosonic star}
\label{section4}
Let us now estimate $\mathcal{S}^{\rm typical}$, given by~\eqref{tabs} or~\eqref{taps} for relativistic bosonic star solutions. First we estimate the maximal mass of a bosonic star~\cite{Liebling:2012fv} from the assumption they are macroscopic quantum states, that is a a very large collection of particles in the same quantum state ($Q\gg 1$).  In this discussion factors of order 1 are dropped and the symbol $\sim$ should be understood as equality up to factors of order 1.

From Heisenberg's uncertainty principle, $\Delta X \Delta P \gtrsim \hbar$; taking $\Delta X\sim R$ as the ``radius" of the bosonic star\footnote{Strictly speaking bosonic stars to not have a radius, since the scalar/Proca field only vanishes at infinity. Nonetheless, due to the exponential decay of the field, one can define a radius, say, as the region enclosing 99\% of the star's mass.}, and $\Delta P\sim mc$ as the maximum momentum of the constituent particles,  one obtains, recalling~\eqref{icw}, 
\begin{equation}
\mu R\gtrsim 1 \ .
\label{radius}
\end{equation}
This formula has a nice interpretation: the radius of a bosonic star is bounded below by the Compton wavelength $1/\mu$ of the constituent particle. 

Saturation of the inequality~\eqref{radius}  corresponds to the minimum radius $R^{\rm min}$ of a bosonic star sourced by a bosonic particle with Compton wavelength $1/\mu$,  
\begin{equation}
\mu R^{\rm min} \sim 1 \ . 
\end{equation}
To make progress, we assume that it is at this minimal radius that the maximal compactness and 
maximal mass of the boson stars is achieved.\footnote{This assumption is corroborated by the analysis of the explicit solutions - see $e.g.$~\cite{Herdeiro:2017fhv,Herdeiro:2019mbz}.}  Assuming this maximal compactness to be comparable to that of a Schwarzchild black hole, then the maximal mass is $GM^{\rm max}\sim R^{\rm min}c^2$ and thus
\begin{equation}
\frac{mM^{\rm max}}{M_{\rm Pl}^2}\sim 1 \ ,
\label{mmass}
\end{equation}
where we introduced the Planck mass $M_{\rm Pl}\equiv \sqrt{\hbar c/G}$. In fact, the explicit computation of the solutions shows that the maximal mass obeys $mM^{\rm max}/M_{\rm Pl}^2=\alpha_{\rm BS}$, where $\alpha_{\rm BS}$ is a model dependent constant of order one. For instance,  $\alpha_{\rm BS}=0.633/1.058$ for the fundamental static scalar/vector mini-bosonic stars~\cite{Herdeiro:2017fhv} and $\alpha_{\rm BS}=1.315/1.125$ for the fundamental spinning scalar/vector mini bosonic stars~\cite{Herdeiro:2019mbz}.\footnote{The term "mini" refers to the model without self-interactions of the bosonic field.} 
Thus~\eqref{mmass} is a very good
estimate.

We can now estimate the number of particles $Q^{\rm max}$ in a bosonic star with maximal mass, assuming that such mass is of the order of the product of the number of particles by the mass of an individual boson: 
\begin{equation}
Q^{\rm max}m\sim M^{\rm max} \ .
\end{equation}
Note that even if the binding energy is a considerable percentage of the total energy, this estimate gives the right order of magnitude. Then
\begin{equation}
Q^{\rm max}\sim \left(\frac{M_{\rm Pl}}{m}\right)^2\sim \left(\frac{M^{\rm max}}{M_{\rm Pl}}\right)^2\sim 10^{76}\left(\frac{M^{\rm max}}{M_\odot}\right)^2 \ .
\label{mq}
\end{equation}
The last estimate is striking: for a bosonic star model that can reach the solar mass, the corresponding bosonic stars with mass $\sim M^{\rm max}$  are made up of $\sim 10^{76}$ particles. Moreover, the number of particles for such stars in models for which $M^{\rm max}$  ranges in the interval~\eqref{massinterval}, is
\begin{equation}
Q^{\rm max} \ \in \ [10^{76},10^{96}]  \ .
\label{qinterval}
\end{equation}

Thus, hypothetical astrophysical bosonic stars are such very large collections of particles in the same quantum state.

\section{Conclusion: a criterion for classicality}
\label{section5}

To provide a physical meaning to the dimensionless term $ \omega  \Delta t$ in~\eqref{tabs} or~\eqref{taps} we observe that if $1/\mu$ provides an estimate of the bosonic star radius (at maximal mass), then 
\begin{equation}
 \mu c\sim \frac{1}{(\Delta t)_{\rm lct}}  \ ,
\label{lct}
\end{equation}
is the inverse  \textit{light crossing time} of the star, denoted  $(\Delta t)_{\rm lct}$. The natural scale for the oscillating frequency of the star $\omega$ is precisely $\omega\sim \mu c$. In fact, the explicit computation of the solutions shows that $\omega/\mu c$ takes values in an interval of the form $]\beta, 1[$, where $\beta$ is model dependent. For instance $\beta\sim 0.767$ for the fundamental, spherical scalar mini-boson stars.

Thus, for the most compact solutions $\mu R^{\rm min} \sim 1$, we have $\omega\sim 1/(\Delta t)_{\rm lct}$ which, together with~\eqref{mq}  recasts~\eqref{tabs} as
\begin{equation}
\frac{\mathcal{S}^{\rm typical}}{\hbar}\sim \left(\frac{M^{\rm max}}{M_{\rm Pl}}\right)^2 \frac{\Delta t}{(\Delta t)_{\rm lct}} \sim   10^{76}\left(\frac{M^{\rm max}}{M_\odot}\right)^2 \frac{\Delta t}{(\Delta t)_{\rm lct}} \ .
\label{stbs}
\end{equation}

Causality implies that for any global physical process of the bosonic star $\Delta t\gtrsim (\Delta t)_{\rm lct}$. Thus, formula~\eqref{stbs} reveals that for bosonic stars  with masses $\sim M^{\rm max}\gg M_{\rm Pl}$ the classical description is legitimate. 
In particular for models where  $M^{\rm max}\gtrsim M_{\odot}$ 
\begin{equation}
\frac{\mathcal{S}^{\rm typical}}{\hbar}\gtrsim  10^{76}\left(\frac{M^{\rm max}}{M_\odot}\right)^2  \gg 1 \ .
\end{equation}
Clearly, this applies for bosonic stars in the mass range~\eqref{massinterval}.

This discussion legitimates the often used purely classical treatment for bosonic stars, as long as the above premisses are obeyed.

%
%

\section*{Acknowledgements}
This work is supported by the Center for Research and Development in Mathematics and Applications (CIDMA) through the Portuguese Foundation for Science and Technology (FCT - Fundaç\~{a}o para a Ci\^{e}ncia e a Tecnologia), references UIDB/04106/2020, UIDP/04106/2020.  We acknowledge support from the projects PTDC/FIS-OUT/28407/2017, CERN/FISPAR/0027/2019 and PTDC/FIS-AST/3041/2020. This work has further been supported by the European Union’s Horizon 2020 research and innovation (RISE) programme H2020-MSCA-RISE-2017 Grant No. FunFiCO-777740.



\end{document}